\documentclass[useAMS,usenatbib]{mn2e}
\pdfoutput=1
\title[Dependence of fragmentation in self-gravitating accretion discs on
small scale structure]{Dependence of fragmentation in self-gravitating accretion discs
on small scale structure}
\author[Young et al.]{
M.D.~Young$^1$, 
and
C.J. Clarke$^1$ \\
$^1$Institute of Astronomy, University of Cambridge, Madingley Road, Cambridge,
CB3 0HA, United Kingdom\\
}
\usepackage{graphicx}
\usepackage{natbib}
\usepackage{url}
\usepackage{amsmath}
\usepackage{amssymb}
\usepackage{caption}
\usepackage{subcaption}
\newcommand{\gv}[1]{\ensuremath{\mbox{\boldmath$ #1 $}}} 
\newcommand{\abs}[1]{\left| #1 \right|} % for absolute value
\newcommand{\grad}[1]{\gv{\nabla} #1} % for gradient
\let\baraccent=\= % rename builtin command \= to \baraccent
\renewcommand{\=}[1]{\stackrel{#1}{=}} % for putting numbers above =

\begin{document}
\date{Written March 2015}
\maketitle
\begin{abstract}
We propose a framework for understanding the fragmentation criterion
for self-gravitating discs which, in contrast to studies
that emphasise the `gravoturbulent' nature of such discs, instead
focuses on the properties of their quasi-regular spiral structures
as derived from simulations. This interpretative framework is shown
to be consistent with existing 2D and 3D numerical studies as well
as with the 2D grid based and SPH simulations conducted here. We propose
two evolutionary paths to fragmentation and argue that
the correct simulation of each of these involves different numerical requirements: i)
collapse on the free-fall time, which requires that the ratio of
cooling time to dynamical time ($\beta$) $< 3$ and ii) quasistatic
collapse on the cooling time at a rate that is sufficiently fast that
fragments are compact enough to withstand disruption when they
encounter spiral features in the disc. We argue that the previous
finding of \cite{PKedge} (in which 2D grid based simulations
demonstrate a numerically converged fragmentation limit of $\beta < 3$
and which we reproduce here with both 2D grid based and 2D SPH simulations)
is a consequence of the fact that such simulations smooth the
gravitational force on a scale of $H$, the scale height of the disc.
Such simulations thus only allow fragmentation via route i) above since
they suppress the quasistatic contraction of fragments on scales $< H$;
the inability of fragments to contract to significantly smaller scales
then renders them susceptible to disruption at the next spiral
arm encounter. On the other hand, 3D simulations (along with 2D simulations
that, with questionable realism, smooth gravity on smaller scales) indeed
show fragmentation at higher $\beta$ via route ii). We derive an analytic
prediction for fragmentation by route ii) based on the requirement
that  fragments can contract sufficiently to withstand disruption by
spiral features, basing this calculation on the properties of spiral
structures derived from simulations. We find that this leads to
a predicted maximum $\beta$ for fragmentation of
 $\sim 12$ , in good agreement with all previous well resolved
3D simulations. We also discuss the necessary numerical requirements on both grid based
and SPH codes if they are to model fragmentation via route ii).
\end{abstract}

\section{Introduction}
\label{sec:intro}

Observations have identified massive planets orbiting at large distances from
their host stars \citep{HR8799}.  Such planets cannot have formed at their
present location via the core-accretion mechanism and migration after
formation at smaller radii is unlikely in some cases \citep{HR8799ecc}.  An
alternative explanation is that such planets formed near their present
locations via direct gravitational collapse \citep{GiantPlanetGravInstab}.
However, the exact disc properties necessary to form planets via the
gravitational instability are still unknown.

The gravitational stability of a disc has typically been understood  in terms
of the toomre $Q$ parameter \citep{toomreQ}, given by
\begin{equation}
  Q = \frac{c_s \kappa}{\pi G \Sigma}
  \label{eq:Qgen}
\end{equation}
where $c_s$ is the speed of sound, $\Sigma$ is the surface density of the disc
and $\kappa$ is the epicyclic frequency ($\Omega$ for a Keplerian disc).
Perturbation analysis of an axisymetric, infinitely thin disc indicates that
$Q>1$ is required for stability.  

When $Q \lesssim 1$ the gravitational instability creates spiral waves. The
shock heating produced by these spiral waves stabilises the disc by increasing
$Q$. This leads to a self-regulated $Q \sim 1$ state in which cooling is 
balanced by spiral wave heating. However, if the gas can cool efficiently, such a 
state is not established and over-densities instead collapse into bound objects.

More than a decade of computer simulations have confirmed that discs with 
$Q \lesssim 1$ are unstable, but can only fragment when cooling is efficient
\citep{Gammie01,BetaCooling,Rice05,MeruBate1,MeruBate2,PKedge}.

Because the evolution of $Q$ is driven most rapidly by changes in 
temperature \citep{DiscReview}, much effort has gone into defining the
cooling rate necessary for discs to fragment.  The simplest approach
parametrises the cooling rate via a constant $\beta$ as
\begin{equation}
  \frac{d u}{d t} = - \frac{u}{\beta\Omega}
  \label{eq:beta_cooling}
\end{equation}
where $u$ is the internal energy per unit mass.

Early studies found that, for equations of state with $P\sim \rho^{\gamma}$
and $\gamma=5/3$,  $\beta > \beta_{crit} \approx 6$ was sufficient to prevent
fragmentation \citep{Rice05}. It was later realised by
\cite{MeruBate1} that this result was not numerically converged, implying 
that fragmentation for slower cooling rates (higher $\beta$) may be possible.
Such non-convergence is unexpected, since even the lowest resolution
simulations resolve the disc scale height $H$, which is expected to be the
most unstable wavelength.

Several possible explanations of this non-convergence have been proposed.
\cite{honHRes,MeruBate2} suggested that unwanted artificial viscosity 
heating (which is resolution dependent) could stabilise discs against fragmentation.  
\cite{RiceCool} suggested that a modification of the
implementation of $\beta$ cooling in SPH could resolve the convergence issue.
\cite{PKedge} found that when starting from commonly used smooth initial
conditions, a boundary layer develops between laminar and 
gravitationally structured flows and can induce fragmentation.

Fundamentally, the non-convergence of the fragmentation boundary can only be
explained if the results are corrupted by numerical effects, or if it is
physically necessary to resolve length scale small than the Jeans' length.
This latter interpretation would be at odds with expectations
based on the dispersion relation for gravitationally unstable discs (Section
\ref{sec:pcollapse}), which implies that the most unstable wavelength is $\sim H$
and that the required amplitude for triggering collapse increases at smaller
and larger spatial scales.  \cite{Hopkins} suggested that rare high amplitude
fluctuations on scale $<H$ can lead to gravitational collapse and used this to
argue for a stochastic fragmentation model that could operate even at very
large values of $\beta$.  Part of the motivation for the present study is thus
to examine whether, in high resolution studies (with up to $30$ resolution
elements across $H$), we see any evidence of this second mode.
In Section \ref{sec:results} of this paper we present a range of new simulations 
examining the fragmentation boundary in 2D using both SPH and FARGO.  We find
that fragmentation is never {\it initiated} on scales $\ll H$ (in contrast to
the hypothesis above), but that these scales become important in modelling the
contraction of an unstable region to a size where it can survive disruption by
spiral shocks.  In Section \ref{sec:spiral} we expand on this argument,
providing estimates based on characterisation of the quasi-regular spiral
shock structure in the simulations.  Finally, in Section \ref{sec:discussion} 
we discuss how this relates to the results presented in the literature and 
its implications for the physical process of fragmentation.

\section{Numerical methods}
\label{sec:model}

\subsection{Disc model}

We initialise all our simulations using a power law in surface density and
temperature and a vertically isothermal Gaussian with scale height $H$.  We
fix the index of the surface density and choose the temperature scaling so
that $Q$ is initially constant.  That is,
\begin{eqnarray}
  \Sigma & = & \frac{M_D}{\pi R_i^2} \left( \frac{R}{R_i} \right)^{-p}
  \frac{\Gamma}{2} \\
  c_s & = & \sqrt{\frac{GM}{R_i}} \left( \frac{R}{R_i}\right)^{-p+3/2}
  \frac{qQ_0 \Gamma}{2} 
  \label{eq:initCond}
\end{eqnarray}
where $\Sigma$ is the surface density, $M$ is the star's mass, $M_D$ the disc 
mass, $R_i$ the inner radius of the disc and $c_s$ is the sound speed of the 
gas. $q=M_D/M$, $Q_0$ is the initial value of $Q$ in the disc, $p$ is a 
parameter to be specified and $\Gamma$ is given by,
\begin{equation}
  \Gamma^{-1} = 
  \begin{cases}
    \log(\xi) \hfill & \text{if $p=2$} \\
    \frac{1}{2-p}(\xi^{2-p}-1) \hfill & \text{if $p\ne 2$} \\
  \end{cases}
  \label{eq:gamma}
\end{equation}
where $\xi=R_o/R_i$ and $R_o$ is the outer radius of the disc.  The disc
aspect ratio $H/R$ is then given by
\begin{equation}
  \frac{H}{R} = \frac{\Gamma Q_0 q}{2} \left( \frac{R}{R_i} \right)^{2-p}
  \label{eq:HonR}
\end{equation}
while the ratio of the resolution scale $h$ to the scale height $H$ is,
\begin{equation}
  \frac{h}{H} = \left( \frac{8\pi}{Nq^2 Q_0^2 \Gamma^3} \right)^{1/d} 
  \left( \frac{R}{R_i} \right)^{\frac{3(p-2)}{d}}
  \label{eq:honH}
\end{equation}
where $d$ is the number of dimensions.  Note that the normalisation of $h/H$
will change by a factor of order unity between different types of code.

We have slightly belaboured this point to emphasise the motivation for our
parameter choices.  The canonical choice in the literature is to choose $p=1$.
Equations \ref{eq:HonR} and \ref{eq:honH} show that this choice results in a
radial gradient in both aspect ratio and resolution, which breaks the otherwise
scale free nature of the problem.  To allow all radii in our simulations to be
treated equally, we instead set $p=2$.  We also choose $q=.2$ and $\xi=5$ to
reduce the computational expense of our simulations.  Following \cite{PKedge}
we use the equation of state $P=K\Sigma^{\gamma}$ with the adiabatic 
exponent $\gamma=5/3$.

Our SPH simulations were performed using a modified version of the popular
code GADGET2 \citep{Gadget2Code}.  The code was modified to include 
artificial conductivity \citep{RiemannSPH}, $\beta$ cooling, particle accretion 
and the correct treatment of softening with variable smoothing lengths
\citep{gravSoftTerms}.  We also implemented the improved artificial viscosity 
method of \cite{Cullen}, which aims to minimise the amount of artificial viscosity 
present away from shock fronts, while still resolving
shocks correctly.  We set the strength of the artificial viscosity (and
conductivity) to the values that produce the best results in test problems
where the correct result is known (e.g., shock tube test, Sedov blast wave,
Kelvin-Helmholtz instability).
For artificial conductivity we use $\alpha_{cond}=1.0$.
For the standard implementation of artificial viscosity, the appropriate
values are $\alpha_{SPH}=1.0$ and $beta_{SPH} = 2.0$ \citep{RiemannSPH}.  
For the artificial viscosity method of \cite{Cullen}, we use $\alpha_{max} = 5.0$, 
$\alpha_{min}=0.0$ and $l=0.05$ \citep{Cullen}.  Note that 
although $\alpha_{max}$ is quite high, in practice the average per-particle value 
is only $\alpha_{SPH} \sim 0.1$ away from shocks.

Grid based simulations were performed using the FARGO code \citep{FARGO}.  
FARGO uses a polar grid, which is logarithmically spaced in the radial direction and
linearly spaced in the azimuthal direction.  The numbers of radial and
azimuthal bins were chosen so that each cell is approximately square (i.e.
$R\Delta \phi \approx \Delta R$).

All of the simulations in this paper are performed in 2D.  This is partially
to allow comparison with FARGO, which only operates in 2D with self-gravity, 
but also to maximise the simulation resolution ($h/H$), which scales as 
$N^{-1/d}$ ($d$ being the number of dimensions).

After initialising each simulation as described above, we run each simulation
for $700 t_{dyn}$ at the inner edge (10 outer rotation periods) with
$\beta=30$ to allow the disc to settle into the $Q \sim 1$ state without
fragmenting.  We then linearly decrease $\beta$ from $30$ to the desired value
over the next $700 t_{dyn}$.  A simulation is deemed to be ``non-fragmenting'' 
if it does not fragment in a further $700 t_{dyn}$.  This procedure follows 
\cite{PKedge} \& \cite{SlowIC} and is designed to prevent fragmentation being 
induced by a boundary layer between regions of smooth and turbulent flow.

\subsection{Gravity in 2D}
\label{sec:grav2d}

The aim of two dimensional simulations is to capture the physics of the three
dimensional system as accurately as possible with a two dimensional
representation.  The treatment of the gravitational force requires particular
care.  In 3D, the gravitational potential is given by Poisson's equation
\begin{equation}
  \grad^2 \Phi = 4\pi G \rho
  \label{eq:poisson}
\end{equation}

In 2D, the code does not have access to the volume density and so we must
obtain an expression in terms of $\Sigma$ and other disc quantities.  How best
to do this depends on the details of the numerical code.  FARGO, which
calculates the gravitational force by directly solving Poisson's equation uses
the expression
\begin{equation}
  \grad^2 \Phi = 4\pi G \Sigma \delta(z-\lambda)
  \label{eq:FARGO_grav}
\end{equation}
where $\lambda$ is a softening factor typically set to some multiple of the
disc scale height $H$.  If $\lambda=0$ then the code behaves as if
$\lambda=h$ where $h$ is the grid spacing.

SPH calculates the gravitational force by summing the contribution from each
particle (typically with the aid of a tree).  To account for the vertical
extent of the disc in 2D, we calculate the gravitational force using a
gravitational potential given by,
\begin{equation}
  \Phi = -G \sum_{b=1}^N m_b \phi(\abs{\ensuremath{\mathbf{r}}-\ensuremath{\mathbf{r_b}}},h)
  \label{eq:SPH_potential}
\end{equation}
where $m_b$ is the mass of each particle $h$ is some softening factor and 
\begin{equation}
  \phi = 
  \begin{cases}
    1/h_G^2 ( \frac{4}{3} q - \frac{6}{5} q^3 + \frac{1}{2} q^4), & 0 \le q <1;\\
    1/h_G^2 ( \frac{8}{3} q - 3 q^2 + \frac{6}{5} q^3-\frac{1}{6} q^4 -
    \frac{1}{15q^2}), & 1 \le q < 2;\\
    1/r^2 & q \ge 2
  \end{cases}
  \label{eq:SPH_grav}
\end{equation}
where $q=r/h_G$ \citep{gravSoftTerms}.
Adopting this form for $\phi$, it follows that
\begin{equation}
  \grad^2 \Phi = 4\pi G \left( \frac{7}{10h_G} \Sigma \right) \approx 4\pi G
  \rho \left( \frac{H}{h_G} \right) 
  \label{eq:SPH_grav_full}
\end{equation}
where $\Sigma$ is estimated using SPH interpolation with kernel length $h_G$.

If we set $h_G = h_{SPH}$ in SPH or $\lambda = 0$ in FARGO, the right hand
side of Equations \ref{eq:FARGO_grav} and \ref{eq:SPH_grav_full} will increase
with resolution.  That is, unless an explicit gravitational softening is
provided in 2D simulations of self-gravitating discs, the gravitational force
will increase with resolution, suggesting that the value of the cooling
time required to suppress fragmentation should increase with resolution.  
On the other hand, if the force is softened on the scale $H$, one would expect
simulations to converge because the suppression of gravitational effects on
small scales means that these scales play no role in the fragmentation
problem.  To test this we ran all our simulations twice, once with no
softening except on the resolution scale and once with softening on the scale
$H$.

\section{Results}
\label{sec:results}

\subsection{SPH simulations}

To measure the fragmentation boundary, we performed a series of SPH simulations 
at different values of $\beta$ using the gradually settled initial conditions described 
in Section \ref{sec:model}. We repeated this experiment with the gravitational force
softened according to $h_G=H$ and $h_G=h_{SPH}$.  All
simulations were classified as ``fragmenting'' or ``non-fragmenting'' as
described in Section \ref{sec:model}.

\begin{figure}
  \begin{center}
    \includegraphics[width=0.5\textwidth]{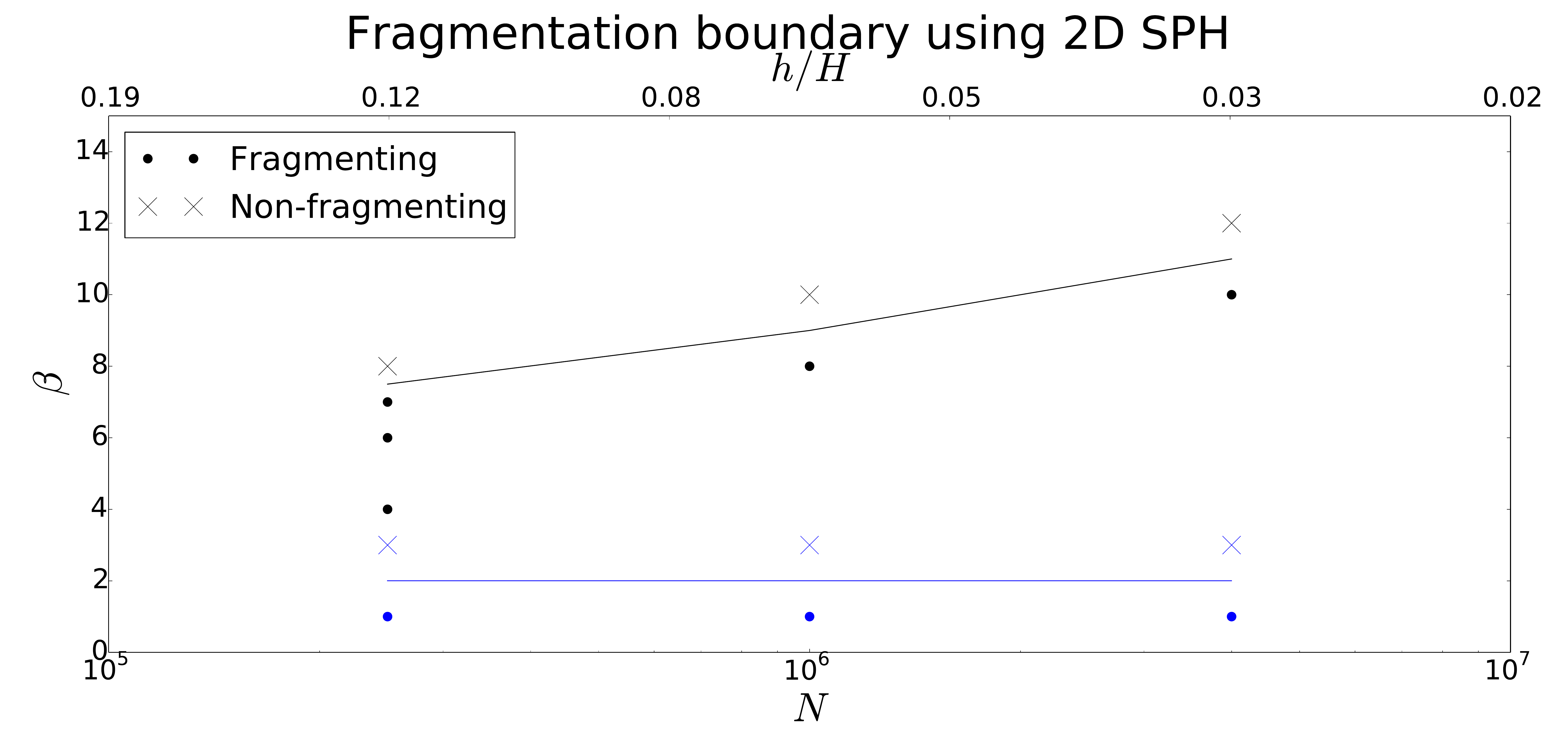}
  \end{center}
  \caption{SPH simulations as a function of cooling time ($\beta$) and
  resolution ($h/H$), with circles denoting fragmenting and crosses
  non-fragmenting simulations, separated by the derived fragmentation boundary
  (solid line).  The lower (blue) symbols are for simulations softened on
  scale $H$ (the disc scale height), while upper symbols are for those
  softened only on the resolution scale ($h$).}
  \label{fig:sphfb}
\end{figure}

Figure \ref{fig:sphfb} shows the results of these simulations with runs
softened on $H$ and $h$ shown in blue and black respectively.  Clearly the
amount of gravitational softening used has a dramatic impact on the
fragmentation boundary.  The simulation softened on $H$ show
convergence to $\beta_{crit} \approx 3$, which is consistent with the value
found by earlier 2D studies \citep{Gammie01} \& \cite{PKedge}.  As well as 
fragmenting at much larger values of $\beta$, those simulations softened only on 
$h$ show no signs of convergence.

\subsection{FARGO simulations}

To better understand the non-convergence of the SPH simulations and to verify
that the SPH results are not an artefact of numerical technique, we ran the
same simulations using the grid code FARGO. Once again, we softened the 
gravitational force on either the resolution scale of the simulation or 
on $H$.  The simulations were initialised and classified into ``fragmenting'' 
and ``non-fragmenting'' in the same way as the SPH simulations.

\begin{figure}
  \begin{center}
    \includegraphics[width=0.5\textwidth]{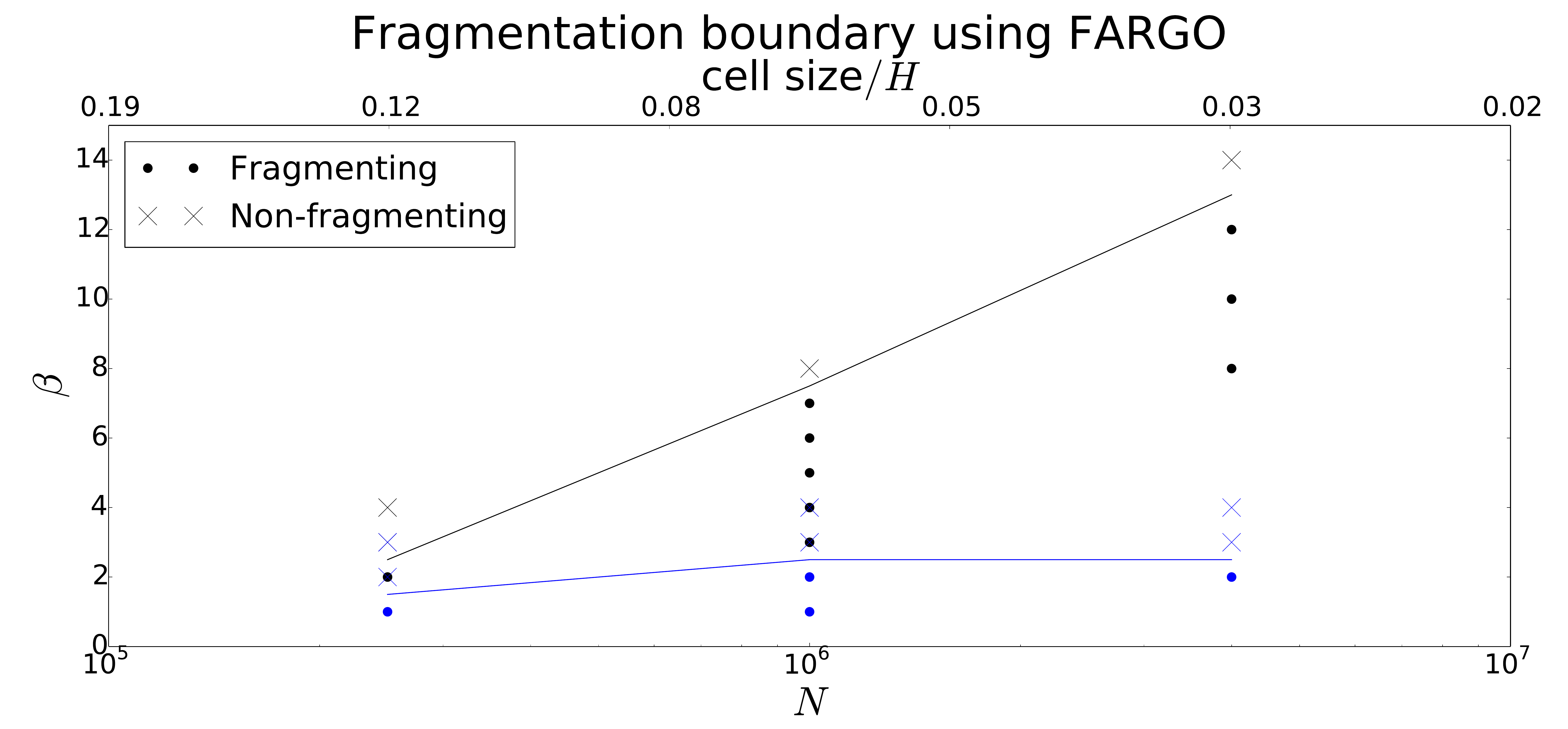}
  \end{center}
  \caption{FARGO simulations as a function of cooling time ($\beta$) and
  resolution (cell size$/H$), with circles denoting fragmenting and crosses
  non-fragmenting simulations, separated by the derived fragmentation boundary
  (solid line).  The lower (blue) symbols are for simulations softened on
  scale $H$ (the disc scale height), while upper symbols are for those
  softened only on the resolution scale.}
  \label{fig:ffb}
\end{figure}

The results are shown in Figure \ref{fig:ffb}.  The runs with gravity softened
on $H$ are shown in blue while those without any extra softening are shown in
black.  The simulations which are softened on the scale of $H$ are numerically
converged to $\beta_{crit} \approx 3$, in good agreement with previous 2D
simulations and the results of our 2D SPH simulations.  Once again, when we
turn off gravitational softening the simulations do not converge
and fragment at similar values of $\beta$ to the SPH simulations at high
resolution.

For those simulations that fragment (in both SPH and FARGO), the fragments form
from structures with length scales $\sim H$.  This remains true even for the
unsoftened simulations that do not converge.  That is, the non-convergence of
our simulations without gravitational softening is not a consequence of 
instabilities triggered on scales significantly smaller than $H$, but of 
how resolution affects the evolution of regions of size $\sim H$ once they
become unstable.

\section{Collapse criteria}
\label{sec:spiral}

The only difference between the converging simulations softened on $H$ and the
non-converging simulations softened on $h$ in Section \ref{sec:results} is on
scales smaller than $H$.  Although unsoftened simulations are physically
incorrect (as they fail to account for the disc's 3D structure), it is at
first sight surprising that processes on scales much less than the Jeans'
length ($H$ for a $Q=1$ disc) should affect the fragmentation process.  

For a disc to fragment, gravitationally unstable over-densities must be able
to collapse into gravitationally bound objects that can resist disruption
\citep{KratterClumps} \& \citep{PKstochastic}.  To determine the role of 
scales smaller than $H$ in fragmentation and estimate when fragmentation is likely, 
we investigate the evolution of gravitationally unstable over-densities in a 
$Q \sim 1$ disc using simplified models for the self-regulated disc structure.

\subsection{Free fall collapse}

The collapse of an over-density into a bound fragment can happen in two
different ways.  If the cooling rate is fast enough, any extra heat generated
by compression of the collapsing over-density is radiated away.  In this
regime, the over-density can never become pressure supported and so collapses
in free-fall.  This will be the case whenever the cooling time is less than
the free-fall collapse time.

For a self-gravitating disc, the free-fall collapse time is given by \cite{KratterClumps}
\begin{equation}
  t_{ff} = \frac{1}{\sqrt{G\rho}} = \sqrt{2\pi Q} t_{dyn}
  \label{eq:tff}
\end{equation}

Therefore, an over density will undergo free-fall collapse whenever
$t_{cool}<t_{ff}$, which is the case whenever
\begin{equation}
  \beta < \sqrt{2\pi Q} \approx 3
  \label{eq:tffbound}
\end{equation}

In this regime, the collapse time $t_{ff} \ll t_{spi}$ (the time scale
on which condensations encounter the next spiral shock: see below)
for all reasonable values of $Q$.  Disc fragmentation is therefore 
inevitable and a self-regulated steady-state cannot form if over-densities 
collapse in free-fall.

\subsection{Pressure supported collapse}
\label{sec:pcollapse}

If $\beta>3$ then gravitationally unstable over-densities become pressure 
supported before they can completely collapse.  Further collapse can only
occur when the extra heat generated by collapse is removed, which happens on
the cooling time scale.  Such pressure supported clumps will only survive if
they can collapse before being disrupted by their environment.

Although the $Q\sim 1$ quasi-equilibrium is constantly changing, it is still
possible to define the average geometric properties of the spiral waves that
are created by the gravitational instability \citep{Cossins1}.  In particular,
we describe the spiral waves using an azimuthal wavenumber $m$ and a radial
wavenumber $k$, which may vary as a function of $R$.  

Let us now consider a patch of disc that has become gravitationally unstable.
On average, the longest it will survive for before encountering a spiral shock is
\begin{equation}
  t_{spi} = \frac{2\pi}{m\abs{\Omega-\Omega_p}} = \frac{2\pi}{m\xi} t_{dyn}
  \label{eq:tspi}
\end{equation}
where $\xi = \abs{\Omega-\Omega_p}/\Omega$ and $\Omega_p$ is the pattern speed of 
the spiral arms.

The core of our argument is that the survival of over-densities (and hence 
the ability of a self-gravitating disc to progress unstable over-densities 
to gravitationally bound fragments) depends on an over-density's ability to
collapse before being disrupted by encountering a spiral shock.  To first order, 
collapse of an over-density takes place on the cooling time scale 
$t_{cool} = \beta t_{dyn}$ and over-densities live for at most $t_{spi}$ 
before encountering a spiral wave.  Therefore, collapse and fragmentation 
requires that $t_{cool} < t_{spi}$, which is true if
\begin{equation}
  \beta < \frac{2\pi}{m\xi}
  \label{eq:betalims}
\end{equation}

Provided that the tight-winding assumption holds ($\abs{kR} \gg m$), it must be 
the case that
\begin{equation}
  M = \frac{m \xi}{kH}
  \label{eq:mach}
\end{equation}
where $M$ is the Mach number of the spiral shock.

Substituting Equation \ref{eq:mach} into Equation \ref{eq:betalims} implies
that
\begin{equation}
  \beta_{crit} = \frac{2\pi}{m\xi} = \frac{2\pi}{MkH}
  \label{eq:bcrit}
\end{equation}

Given a dispersion relation, it is possible to re-write $kH$ in
terms of the Toomre parameter $Q$ and the Mach number $M$.  If we assume an
infinitely thin, tightly would disc, the standard dispersion relation can
be re-written as \citep{BandT},
\begin{equation}
  m^2\xi^2 = H^2 k^2 -\frac{2H\abs{k}}{Q} + 1
  \label{eq:disprel}
\end{equation}
substituting in Equation \ref{eq:mach} and solving for $kH$ gives
\begin{equation}
  kH = \frac{-1 + \sqrt{1+Q^2(M^2-1)}}{Q(M^2-1)} \approx \frac{Q}{2}
  \label{eq:dispsolv}
\end{equation}
where the last equality holds when $M - 1 \ll 1$.

Substituting Equation \ref{eq:dispsolv} into Equation \ref{eq:bcrit} gives
\begin{equation}
  \beta_{crit} = \frac{4\pi}{MQ}
  \label{eq:bcritQ}
\end{equation}
where $Q$ is to be determined numerically.

Although it may seem that $Q$ is unconstrained in this model, it is
actually equivalent to setting the most unstable wavelength.  For example, if
the most unstable wavelength is similar to that of the axisymmetric disc
(where $kH=1/Q$), then it must be the case that $Q = \sqrt{2}$.  Numerical
simulations typically find that $1<Q<\sqrt{2}$.  Together with the expectation
that $M \approx 1$, this leads us to predict that $\beta_{crit} \approx 9-13$
for well resolved simulations.

Equations \ref{eq:bcrit} \& \ref{eq:bcritQ}  are important as they provide a simple 
analytic estimate of $\beta_{crit}$ (the first, to our knowledge).  That said, the
arguments made in reaching this estimate are all ``first-order'' and so we
expect Equation \ref{eq:bcritQ} to be accurate only up to a factor of order
unity.  Furthermore, Equation \ref{eq:disprel} assumes an infinitely thin
disc.  If we use a dispersion relation that accounts for disc thickness (see
Equation 31 of \cite{Cossins1}), the resulting estimate of
$\beta_{crit}$ is reduced slightly.  More explicitly,
\begin{equation}
  \beta_{crit} = \frac{4\pi}{MQ} (1-\frac{Q}{2})
  \label{eq:bcritQthick}
\end{equation}
and the most unstable wavelength changes so that we expect $Q=1$.

\subsection{Resolution requirements}

To understand the resolution requirements of resolving the collapse process
described in the previous section, we 
need to know how ``collapsed'' a clump needs to be in order to survive a 
collision with a spiral arm.  The average amount of energy per unit mass lost
to cooling between spiral wave encounters is roughly $u t_{spi}/t_{cool}$.  If 
shock heating balances cooling (as it must in the $Q \sim 1$ state), 
material will gain
\begin{equation}
  \Delta u = u t_{spi}/t_{cool} = \frac{2\pi}{m\xi \beta} u
  \label{eq:delu}
\end{equation}
in internal energy per encounter with a spiral.  

In order for a clump to survive a collision with a spiral arm, it must still
be smaller than its initial size, once this extra $\Delta u$ energy
has been deposited.  That is, collapse begins with a patch of size $\sim H$,
which then contracts until it is hit by a spiral wave.  This spiral wave
deposits energy in the clump, causing it to expand.  If the expanded size of
the post-shock clump is greater than the size when collapse began, the clump
will no longer be unstable and will not survive.  

The binding energy of the clump with size $x$, per unit mass, is given by,
\begin{equation}
  e_{bind} = \frac{2GM_c}{3x}
  \label{eq:ebind}
\end{equation}
and so the post-shock clump size, $x_2$, is given by,
\begin{equation}
  \frac{2GM_c}{3x_2} = \frac{2GM_c}{3x}-\frac{2\pi u}{m\xi \beta}
  \label{eq:csurv}
\end{equation}
If we require that $x_2 < H$, this implies that
\begin{equation}
  \frac{x}{H} < \left( 1 + \frac{3\pi Q}{m\xi \gamma (\gamma-1) \beta}
  \right)^{-1}
  \label{eq:csurvfin}
\end{equation}

For reasonable values of $\gamma=5/3.$, $Q=1$ and $\beta=6$, this yields the
requirement that $x/H < .25$.  Furthermore, if we derive a simple expression for 
$x(t)$ (see \cite{KratterClumps} for example), substitute $x(t_{spi})$ for 
$x$ using Equation \ref{eq:tspi} and solve for the $\beta$ (for the same values of
$\gamma$ and $Q$), we find that fragmentation occurs when
\begin{equation}
  \beta < \frac{2.2\pi}{m\xi}
  \label{eq:numsolv}
\end{equation}
a requirement that is remarkably similar to Equation \ref{eq:bcrit}, which we
obtained from simple time scale arguments.  The fact that Equations
\ref{eq:numsolv} and \ref{eq:bcrit} are so similar means that requiring
$t_{cool} < t_{spi}$ ensures that the clump has contracted sufficiently to
satisfy Equation \ref{eq:csurv} and hence survive.

However, the derivation of Equation \ref{eq:csurvfin} has required a number of
simplifying assumptions regarding the non-linear evolution of the unstable
over-density.  The true utility of Equation \ref{eq:csurvfin} is not in
providing a more precise constraint on $\beta_{crit}$ than Equation
\ref{eq:bcrit} (both are subject to an uncertainty of order a factor of 2), but
in understanding what physical scales and processes need to be resolved in
order to accurately model the non-linear evolution of the clump and the
formation of fragments.

More specifically, we find that to survive a collision with a spiral wave, a
fragment must have $x/H \ll 1$.  The exact value of $x/H$ required depends on
$\beta$ and like all other arguments in this section is 
uncertain to within a factor of order unity.  Nevertheless, this reasoning
shows why resolving scales smaller than $H$ is necessary to accurately model
the physics of disc fragmentation.  It is necessary to resolve the
physical scale of the initial instability \emph{and} the scale to which that
instability must shrink in order to survive collisions with spiral waves.

\section{Discussion}
\label{sec:discussion}

Most simulations that have investigated the fragmentation boundary in 2D have
found a numerically converged fragmentation boundary of $\beta_{crit}
\approx 3$ \citep{Gammie01,PKedge}.  These 2D simulations softened their
gravitational force on $\sim H$\footnote{\cite{Gammie01} 
does not use an explicit gravitational softening, but is forced to omit small 
wavelength modes in his calculation of the gravitational force 
(which uses a Fourier transform of Poission's equation).  Discarding these 
modes is equivalent to a gravitational softening on the scale $\sim 0.3H$.}, 
similar to our simulations in Section
\ref{sec:results}, which also found $\beta_{crit} \approx 3$.
The only major exception to this is \cite{MeruBate2}, who used a very weak
gravitational softening of $3\times 10^{-4} H$ and obtained a much higher
fragmentation boundary and much weaker numerical convergence.  
The fragmentation boundary also appears to be independent of $\gamma$ in softened
2D simulations, with simulations using $\gamma=2$ \citep{Gammie01} and
$\gamma=5/3$ (\cite{PKedge}, Section \ref{sec:results}) all yielding
$\beta_{crit} \approx 3$.

In contrast to this, all 3D simulations performed to date fragment at
well above $\beta=3$ \citep{Rice05,RiceCool,RiceCool2,Cossins1,MeruBate1,MeruBate2}.
The fragmentation boundary in these simulations also shows differing degrees 
of numerical convergence \citep{MeruBate1,MeruBate2,RiceCool2} and have been
convincingly shown to depend on $\gamma$ \citep{Rice05}.

A consequence of softening gravity on the scale $H$ is that the force between
mass elements separated by less than $H$ goes to zero (to first order).  This
suppression of the gravitational force on small scales means that any clump
that becomes pressure supported on a scale $l<H$ lacks the inwards
gravitational pull necessary to drive it to contract further. In effect,
softening of small scale gravitational forces causes pressure-supported
collapse to stall.  Because of this, fragmentation via pressure-supported
clumps is suppressed in 2D simulations with gravitational softening.

It is only when $t_{cool}<t_{ff}$ that pressure is never able to impede collapse 
and simulations softened on the scale $H$ can fragment.  In this regime, softening 
on $H$ cannot impede collapse since the clumps enter the softened regime in free 
fall and so it is not necessary that they
experience continued inward acceleration to drive collapse.  Since $t_{ff}
\approx 3 t_{dyn}$ this occurs whenever $\beta < 3$, in excellent agreement
with the converged fragmentation boundary identified in our simulations and in
the simulations of \cite{Gammie01} \& \cite{PKedge}.  If the amount of gravitational 
softening is reduced, the fragmentation boundary will increase as the scale at which
pressure supported collapse will be stalled will decrease \citep{MeruBate2}.

If gravitational softening is set by the resolution scale, then the gravitational 
force will be given by,
\begin{equation}
  \grad^2 \Phi = 4\pi G (\Sigma/h)
  \label{eq:bad_grav}
\end{equation}
where $h \sim N^{-1/2}$.  That is, the gravitational force will continue to
increase with resolution.  This increased gravitational pull will make
fragmentation easier as the resolution increases, as is demonstrated in Figures
\ref{fig:sphfb} and \ref{fig:ffb}.  Even if a converged value of the
fragmentation boundary can be reached with gravity softened on the resolution
scale, its physical interpretation is unclear \citep{grav2d}.

\cite{Rice05} showed that the fragmentation boundary varied as
$(\gamma(\gamma-1))^{-1}$ in 3D simulations.  However, the softened 2D simulations 
of \cite{Gammie01}, \cite{PKedge} and those performed here all agree on the 
fragmentation boundary, despite using significantly different values of
$\gamma$.  This finding is also consistent with different types of
simulations (softened 2D, realistic 3D), probing different modes of
fragmentation (free-fall collapse, pressure supported collapse).  
Softened 2D simulations are only able to probe fragmentation
by free-fall collapse, which does not directly depend on $\gamma$ (since the
free-fall time is independent of $\gamma$).  3D simulations also permit
fragmentation via pressure supported collapse, which does depend on $\gamma$
(see Equation \ref{eq:csurvfin}).

It is possible that a more sophisticated model for approximating the gravity
in two dimensions, such as the technique proposed by \cite{grav2d}, would 
reduce the fragmentation suppressing effects of gravitational softening.
However, it is inevitable that some differences with the full 3D solution to
Poisson's equation will remain.  In particular, models that assume a vertical
structure for the disc are likely to be least accurate at modelling the
quasi-spherical over-densities that must be modelled for fragmentation. 
As some gravitational
softening is necessary to approximate the three dimensional force in two
dimensions, any two dimensional simulation must either fail to model the
gravitational force 100\% correctly or suppress fragmentation via the 
collapse of pressure supported clumps.

Three dimensional simulations do not suffer from this limitation and so are
able to model both pressure supported and free-fall collapse.  However, there
remains some disagreement as to the converged value of the fragmentation
boundary in 3D \citep{MeruBate1,MeruBate2,PKedge,honHRes,RiceCool,RiceCool2}.  
The model for pressure supported collapse developed in
Section \ref{sec:spiral} estimates $\beta_{crit} \sim 12$.  Given that we do not
expect this estimate to be more accurate than a factor of $2$, this does not
allow us to discriminate between the $\beta_{crit} \sim 8$ favoured by
\cite{RiceCool2} and the $\beta_{crit} = 20-30$ of \cite{MeruBate2}.  However,
we argue in Section \ref{sec:stoch} that the quasi-regular nature of the spiral
structure in self-gravitating discs lead us to expect that fragmentation
should not be possible at value of $\beta$ significantly higher than the
latter range ($\beta_{crit} = 20-30$), even allowing for the possibility of
stochastic fragmentation.

\subsection{Numerical technique specific issues}

A fixed grid, such as the one used by FARGO, is only able to follow a
contracting clump down to the resolution scale of the simulation.  Given this,
care must be taken that the resolution scale is much smaller than the smallest
scale of physical importance.  We have shown in Section \ref{sec:spiral} that 
this length scale can be significantly smaller than the disc scale height $H$.  
Failure to resolve small enough scales will suppress fragmentation by preventing 
collapse from proceeding to the point where it can resist disruption by interaction
with spiral arms.

By contrast, the adaptive nature of SPH ensures that so long as the resolution
is high enough to resolve the initial instability ($h/H < 1$), it will
continue to resolve the clump as it contracts.  However, modelling the
collapse of pressure supported clumps is challenging in SPH because of the
effect of particle noise.  Any random ``thermal'' motions acquired by SPH
particles will supply an additional pressure that will prevent collapse.
Unlike pressure derived from internal energy, this pressure due to particle 
noise can not be removed by cooling.  

Failure to provide adequate artificial viscosity at shock fronts is a well 
known source of particle noise in SPH. Strong shear, as is present in 
differentially rotating discs, can also lead to particle noise.  Consistent 
with this interpretation, \cite{MeruBate2} found
that lowering the strength of the artificial viscosity creates particle noise 
and inhibits fragmentation.  It is also worth noting
that there is abundant evidence from tests with analytic solutions that 
the suppression of particle noise at shocks also requires a rather large value 
of $\alpha_{SPH}$ ($>0.7$) \cite{properAV1,properAV2,properAV3}.  This
is significantly larger than the value of $\alpha_{SPH}=0.1$ this is commonly
employed in disc fragmentation calculations 
\citep{Rice05,Cossins1,MeruBate1,DataPaper,RiceCool2}.  Unfortunately, such a
high value of $\alpha_{SPH}$ produces significant heating away from shock
fronts unless the resolution is very high and this can suppress fragmentation,
as has been discussed extensively in the literature 
\citep{MeruBate1,MeruBate2,honHRes,RiceCool2}.

Therefore, special care must be taken to ensure that shock-fronts are well
modelled and particle noise suppressed in order to accurately model
fragmentation.  Unfortunately, doing so greatly increases the number of
particles required to prevent excessive artificial viscosity heating away from
shock-fronts (although this computational cost can be reduced by using modern
SPH techniques such as improved artificial viscosity triggers \citep{Cullen}
and kernels with increased neighbour number \citep{Wend_kern}).

\subsection{Stochastic fragmentation}
\label{sec:stoch}

It was shown by \cite{PKstochastic} that even when simulations settle into the
$Q \sim 1$ state without fragmenting immediately, they can still fragment
stochastically over the lifetime of the disc. \cite{PKstochastic} argued
that this occurred when a clump ``got lucky'' and survived for long enough to
resist disruption and form a fragment.  This process can also be understood
within the framework of quasi-regular, but intermittent, spiral structure.

The expression which relates the geometry of the spiral waves to the
fragmentation boundary, $m\xi$, describes the {\em average} geometry of the
disc.  The spiral arms present at any particular time will fluctuate about
this average geometry.  When $\beta > \beta_{crit}$, the average time between
spiral wave encounters is shorter than the collapse time of an over-density.
As such, the disc is resistant to fragmentation on average.  However, if the
gravitational instability happens to be triggered in a part of the disc with
the right random fluctuations in spiral geometry, it can survive longer than
would be expected from the mean geometry and collapse enough to fragment.

Precise numerical studies are required to quantify the exact likelihood that
this random process will lead to fragmentation, which is beyond the scope of
the present study.

\subsection{General implications}

This paper has argued that two dimensional simulations cannot be
trusted to model the fragmentation process correctly.  Nonetheless, we are
now in a position to provide some important upper limits on when
fragmentation can occur.

We argue that fragmentation is impossible if the initial instability
cannot contract significantly before it encounters a spiral arm.  Since
collapse occurs on the cooling time scale, this occurs whenever
\begin{equation}
  t_{cool} \ge \frac{2\pi}{m\xi} t_{dyn}
\end{equation}
Furthermore, $m\xi$ can be related to $Q$ and the Mach number via the dispersion
relation to give,
\begin{equation}
  t_{cool} \ge \frac{4\pi}{MQ} t_{dyn} \approx 12 t_{dyn}
  \label{eq:beta_upper}
\end{equation}

Equation \ref{eq:beta_upper} is uncertain to within a factor of a few.  Nevertheless, 
we can confidently say that our model predicts that fragmentation will not occur 
whenever $\beta$ is greater than a few tens.  

There is some possibility that stochastic fragmentation can lead to
fragmentation at higher cooling times.  Over the self-gravitating lifetime of
a disc, this may shift the ``effective fragmentation boundary'' a factor of
two or so higher (although detailed numerical studies are needed to confirm
this).  

The opacity regime in the outer parts of protoplanetary discs leads to 
$\beta \propto R^{-9/2}$ \citep{CathieChemFrag,Cossins2,PKstochastic}.  
Since $\beta$ is such a strong function of radius, a factor of a few 
uncertainty in $\beta_{crit}$ makes 
little practical difference to planet formation 
theory.  That is, the upper bounds provided by Equation \ref{eq:beta_upper} 
already place the allowable radii for fragmentation with sufficient precision.

\section{Conclusions}

In this paper we have presented a new interpretation of the fragmentation
process in discs, which emphasises the role of the quasi-regular spiral
structure.  In this interpretation, a disc can only fragment when an
over-density is able to contract to a size small enough to survive being disrupted when
it encounters a spiral arm.  We show that this requirement means that
simulations of fragmenting discs need to resolve both the initial instability
at scale $H$ and the subsequent collapse of the resulting over-densities to
a size $x \ll H$.

We have further shown that the fragmentation boundary obtained in 2D
simulations depends strongly on the type of gravitational softening employed.
We have found that although gravitational softening on the scale $H$ is
necessary to account for the vertical structure of the disc and obtain a
converged fragmentation boundary, it also suppresses fragmentation by
preventing the collapse of pressure supported clumps.  As such, it is perhaps 
unlikely that any 2D model of a self-gravitating disc will be able to capture 
all the processes necessary to model disc fragmentation in its entirety.

The necessity of modelling the contraction of a clump from instability through
to disruption resistant fragment also leads to additional numerical
requirements for simulations of fragmenting discs.  For codes employing a
fixed-grid, it imposes the requirement that scales $x \ll H$ be resolved so
that clumps can collapse enough to resist disruption by spiral waves.  For SPH
calculations, where this resolution requirement is automatically satisfied if
$H$ is well resolved, it leads to the requirement that particle jitter be
eliminated.  To achieve this, the artificial viscosity must be calibrated to
produce the correct results in test problems where the answer is known.
Additionally, the widely discussed requirement that artificial viscosity heating be
small compared to the imposed cooling must also be satisfied.

Unfortunately, satisfying all these requirements (3D simulations, no
particle-jitter, minimal artificial viscosity heating) requires a very high
particle number.  Even for the parameters used in this paper, which were
chosen to maximise computational efficiency, achieving an artificial viscosity
heating less than 10\% of the cooling rate when $\beta=20$ with the standard
fixed artificial viscosity method ($\alpha_{SPH}=1.0$) requires of order 
$400$ million particles in 3D.

Despite these severe numerical limitations, the time scale arguments presented
in this paper allow us to predict that $\beta_{crit}$ can be at most a few
tens.  Given the strong scaling of $\beta$ with radius in realistic
protoplanetary discs, a more precise constraint on $\beta_{crit}$ is unlikely
to be of physical importance for disc fragmentation as a mechanism for giant
planet formation.

\section{Materials \& Methods}
\label{sec:materials}

In the interests of reproducibility and transparency, all code and data used
in performing this work have been made freely available online at
\url{https://bitbucket.org/constantAmateur/discfragmentation}.

\section{Acknowledgements}
\label{sec:ack}

This work benefited greatly from discussions with Giuseppe Lodato, Deborah Sijacki,
Richard Nelson, Farzana Meru, Richard Booth \& Daniel Price on issues of
both physics and numerics.  We would like to thank Richard Booth \& Sijme-Jan 
Paardekooper for providing valuable feedback on the manuscript.  Sijme-Jan
Paardekooper deserves special thanks for having provided advice and
feedback throughout the course of this project.  We thank the referee
for comments which improved the paper.

Matthew Young gratefully acknowledges the support of a Poynton Cambridge
Australia Scholarship. The simulations in this work were performed using HPC
resources allocated through DiRAC project DP022.  This work has been supported 
by the DISCSIM project, grant agreement 341137 funded by the European 
Research Council under ERC-2013-ADG.

\bibliographystyle{mn2e}
\bibliography{references}

\end{document}